\documentclass[intlimits,twoside,a4paper]{article}

\usepackage{amsmath,amssymb}
\usepackage{graphicx}

\usepackage[T2A]{fontenc}
\usepackage[cp1251]{inputenc}

\usepackage[eqsecnum]{cmpj2}

\usepackage{color}

\issue{2016}{19}{1}{13607}
\doinumber{10.5488/CMP.19.13607}

\title[Profiles of eletrostatic potential]%
{Profiles of electrostatic potential across the water-vapor, ice-vapor
and ice-water interfaces }%
\author[T. Bryk, A.D.J. Haymet]{T. Bryk\refaddr{label1,label2},
        A.D.J. Haymet\refaddr{label3}}
\addresses{
\addr{label1} Institute for Condensed Matter Physics of the
National Academy of Sciences of Ukraine, \\
1~Svientsitskii St., 79011 Lviv, Ukraine
\addr{label2} Institute of Applied Mathematics and
Fundamental Sciences, Lviv Polytechnic National University, \\ 12 Bandera St.,  79013
Lviv, Ukraine
\addr{label3}
Scripps Institution of Oceanography, UC San Diego, San Diego,
California 92093, USA
}

\authorcopyright{T. Bryk, A.D.J. Haymet, 2016}
\date{Received November 22, 2015, in final form January 4, 2016}

\begin{document}
\maketitle

\begin{abstract}
Ice-water, water-vapor interfaces and ice surface are studied by molecular dynamics
simulations with the SPC/E model of water molecules having the purpose to estimate
the profiles of electrostatic potential across the interfaces. We have proposed a methodology
for calculating the profiles of electrostatic potential based on a trial particle,
which showed good agreement for the case of electrostatic potential profile of the
water-vapor interface of TIP4P model calculated in another way. The measured profile of
electrostatic potential for the pure ice-water interface  decreases towards the
liquid bulk region, which is in agreement with simulations of preferential direction
of motion of Li$^{+}$ and F$^{-}$ solute ions at the liquid side of the ice-water interface.
These results are discussed in connection with the Workman-Reynolds effect.

\pacs 61.20.Qg, 05.20.-y, 82.45.Gj
\end{abstract}


\section {Introduction}
 Fundamental presence of water, ice and their ice-water, ice-vapor(air) and
water-vapor(air) interfaces in nature defines huge interest to exploration of their
structural and dynamic properties \cite{Howe,Lai92}. Microscopic structure and
dynamics of the interfaces
can be studied by atomistic molecular dynamics (MD) computer simulations using either models
of water molecules with effective interactions or  on {\it ab initio}
level with
explicit account for electron subsystem. Among the different ice and water interfaces,
the ice-water interface has been studied much less, although by the date the issues of stability
of the water-ice interface \cite{Kar87,Bry04,Yoo11}, its width \cite{Nada95,Baez95,Hay01,Bry02} and
interfacial free energy \cite{Hay05,Han08,Pir11,Dav12} are quite well elaborated.

Not only the pure interfaces are of great interest~--- even more interesting and fascinating
problem is the behavior of different molecular and atomic solutes in the interfacial regions.
One can mention a problem of the effect of solutes on the surface tension of aqueous
solutions \cite{Hey10,Lev09}, a problem of ionic transfer across aqueous and ice interfaces
\cite{Wic08,Lee12,Son14,Son15},
the effects of salt on premelting the surface layers of ice \cite{Das06,Kim08,Bry04b}, and the
Workman-Reynolds effect \cite{Wor50,Gro65,Car80,Car83,Bro91,Wil08a,Wil08b} observed during the freezing
of dilute aqueous
solutions. The Workman-Reynolds effect of the emergence of freezing potential between the
solid and liquid phases is connected with a charge separation occuring at a growing ice surface
(moving ice-water interface). For simple aqueous solutions of NaCl, NaF, NaI, LiCl, etc.,  the
Workman-Reynolds experiments gave evidence of positive potential of liquid with respect to ice,
i.e., the charge separation with an excess of positive ions on the liquid side of the ice-aqueous
solution interface. By date it is not clear what causes the observed charge separation at
the ice-water interface.

Computer simulations support the observations of the charge separation occuring in Workman-Rey\-nolds effect.
Following a molecular dynamics study of the tendencies in ionic solute motion close to
the water/di\-chlo\-ro\-ethane interface \cite{Ben93}, one may use the same methodology
and place positive/negative ions on the liquid side of the ice-water interface. Upon
proper equilibration, one can observe the preferable direction of motion of the ions
towards/outwards the interface. Such
a study with Na$^{+}$ and Cl$^{-}$ solute ions at the liquid side of ice-water interface
was reported in \cite{Hay05}. The positive sodium ions were moving towards the bulk water
region of the two-phase system, while the negative Cl$^{-}$ ions were trying to penetrate
deeper into the interface heading towards the ice bulk region.
These results for behaviour of positive/negative ions at the
ice-water interface imply that the preferential direction of motion of solute ions
 can be caused by a difference in electrostatic potential between the
solid and liquid sides of the pure interface. Therefore, the aim of this study was
to check if the same effect is observed for the ions at the ice-water interface
other than those reported in \cite{Hay05}, and to estimate
the profiles of
electrostatic potential across three different interfaces: the water-vapor, ice-vapor
and ice-water ones~--- and then compare them. The profile of electrostatic potential has already been
calculated for water-vapor interface \cite{Wil88,Wil89}, while for ice surface
there are experimental estimates of the surface potential, as well as there were reported
the profiles of the potential of the mean force for positive/negative ions \cite{Hay05,Bry04b}.
The rest of the paper is organized as follows: in the next section we supply the details of our
molecular dynamics simulations as well as the methodology of  calculations of the profile of
electrostatic potential will be explained. In section~\ref{sec3}, we report the results for the
profiles of electrostatic potential for the water-vapor, ice-vapor
and ice-water interfaces and  discuss our findings. The last section contains a conclusion of this study.

\section{Methodology of calculations}
\label{sec2}

Molecular dynamic simulations for the three two-phase systems (containing water-vapor, ice-vapor
and ice-water interfaces) were performed with the rigid SPC/E model \cite{Jor83} of water molecules.
The simulated model ice-water system consisted of 2304 water molecules, ice-vapor and water-vapor
systems~--- each of 1344 molecules. The average temperature in simulations of ice-water and ice-vapor
systems was 225~K, i.e., nearly at the melting point for ice of the SPC/E model \cite{Bry04,San04},
and for the water-vapor system the temperature was 298~K. For the case of ice-water and ice-vapor
interfaces, we studied only the basal face of the interfaces, although the other orientations
of ice $I_h$ form ice-water interfaces with a bit smaller 10--90 widths than the basal face
(see \cite{Bry02}), however, this should not change the general tendency of the behavior of ions
 at the inerface. Calculations of electrostatic profiles for the prism, ($20\bar{2}1$) and
($2\bar{1}\bar{1}0$) interfaces will be reported elsewhere.

The preparation of the interfaces for simulations at ambient pressure was described in detail in \cite{Hay05}. We used the same sequence of ($NPT$), ($NP_zAT$) and ($NVT$) ensembles in order to first prepare
the bulk ice and the bulk water systems at the same ambient pressure, and then to bring water and ice into
contact having the same area $A$ and equilibrate the two-phase system by fluctuating only
the $z$-box length at  $z$-component of pressure tensor fixed to 1~bar. The size of equilibrated ice-water
system of 2304 SPC/E water molecules was $26.8253~\text{\AA} \times 30.9903~\text{\AA} \times 85.3827~\text{\AA}$.
The ultimate production runs of simulations with the purpose of calculating the electrostatic
potentials were performed
in $NVT$ ensemble with Nose-Hoover thermostats. The electrostatics in simulations was
treated by 3D Ewald method in all three cases of different interfaces.
The cut-off distance for short-range part of
potentials was 10~{\AA}. All the simulations were performed by DL\_POLY package \cite{DL}.

We also performed  $\sim  1.1-1.2$ nanosecond-long simulations of the behaviour of
single Li$^{+}$ and F$^{-}$ ions on the liquid side of the basal face of ice-water interface at $T=225$~K. For ion interaction with the oxygens of SPC/E water molecules we used  the following
parameters for short-range Lennard-Jones potential: $\varepsilon_\text{LiO}=0.160026$~kcal/mol,
$\sigma_\text{LiO} =2.337$~{\AA} and $\varepsilon_\text{FO}=0.167144$~kcal/mol,
$\sigma_\text{FO}=3.143$~{\AA}. The ions at initial positions were slowly grown in at a fixed $z$-position
first as a neutral particle with an increasing size every 500 timesteps  (up to the mentioned
Lennard-Jones parameter $\sigma_\text{LiO}$), and then gradually increasing its charge
by $\pm 0.1$ every 500 steps \cite{Hay05}. This procedure allowed us to grow in the solute ions
without any structural damages for the ice-water interface.

Calculations of the electrostatic potential across the simulated two-phase system can be
performed by means of the estimated charge-density profiles using their Fourier-transforms
to solve the Poisson equation. For the case of ice-water and ice-vapor interfaces, the
charge-density profiles contain positive and negative sharp peaks \cite{Bry02} in the solid
phase which makes a direct application of the standard fast Fourier-transform programs problematic.
In fact, sharp peaks in the charge-density profiles come from the locations of the effective
charges of oxygens and hydrogens in the atomic planes. However, the solute ions never reach
the locations of these point charges on oxygens and hydrogens. Therefore, we used a simple
methodology which is based on a trial neutral particle which moves in a plane with constrained
$z$-coordinate and the single-particle electrostatic energy is calculated every few steps at the
position of the trial particle (as if it had a charge $+1$). The trial particle interacts
via soft-core potential $\sim (\sigma/r)^{12}$ with the oxygens and hydrogens and, therefore,
it practically
moves freely in the atomic plane with constrained $z$ avoiding only the locations of the point
charges on oxygens and hydrogens due to soft-core repulsion.
In that case (having in mind that we can apply the Ewald
methodology to calculate the single-particle electrostatic energy as if the trial particle
had the charge $+1$) the electrostatic potential $V(z)$ is simply the average of
single-particle electrostatic (Ewald) energies over configurations with the same $z$-coordinate
of the trial particle minus the Ewald self-interaction.
Note, that in order to avoid a shift of the interfaces in a system with a
constrained particle, the motion of the simulation cell is constrained by removing the $z$-component
of the velocity at every step during the molecular dynamics simulations, similarly as it was
proposed in \cite{Dan02a,Dan02b} and was used earlier in our simulations reported in \cite{Hay05}.

The $z$-coordinate of trial particle was changing with a step of $0.5$~{\AA}. In order to grow in
the trial particle at a new $z$-position, we slowly increased the effective parameter $\sigma$ every
500 time steps. After several thousand steps, the trial particle was grown in at a new $z$-position
practically without any effect on the structure of the interface.

\section{Results and discussion}
\label{sec3}

\begin{figure}[!b]
\centerline{
\includegraphics[width=0.5\textwidth]{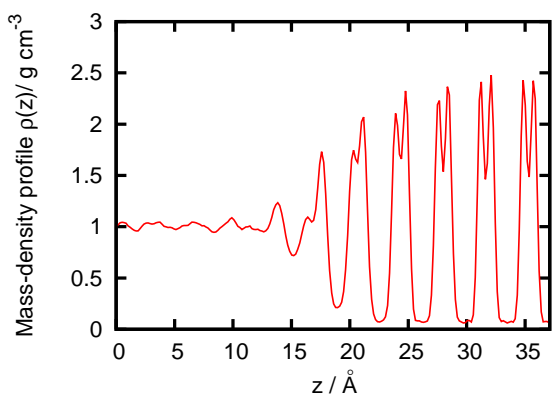}
}
\caption{(Color online) Mass-density profile of the basal face ice-water interface
obtained from simulations with the SPC/E model at $T=225$~K.}
\label{profile}
\end{figure}

\begin{figure}[!t]
\centerline{
\includegraphics[width=0.5\textwidth]{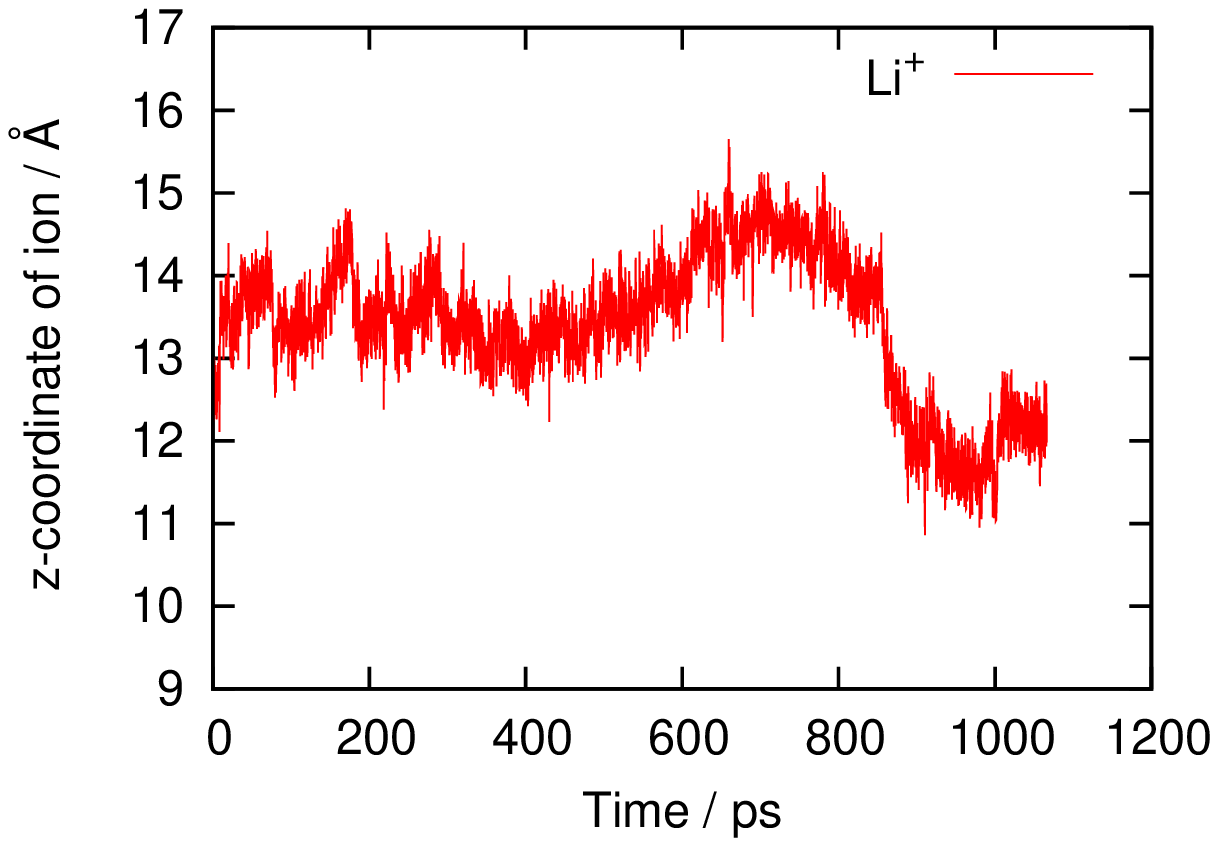}
\includegraphics[width=0.5\textwidth]{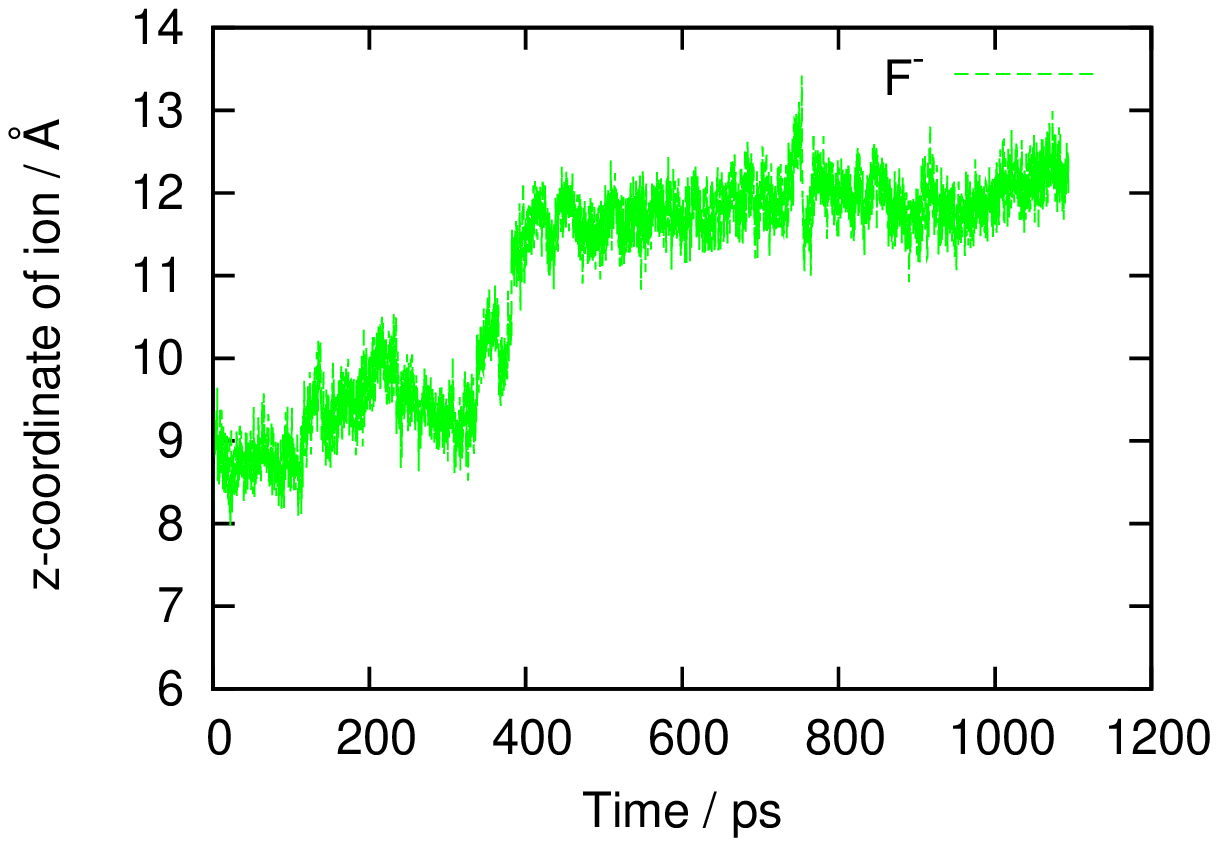}
}
\caption{(Color online) Evolution of $z$-coordinate of Li$^+$ and F$^{-}$ ions
initially placed and equilibrated on the liquid side of the ice-water
interface. The values of $z$-coordinate correspond to the mass-density profile
shown in figure~\ref{profile}.
} \label{ions}
\end{figure}

Although it is known that
the solute ions are expelled from the ice bulk phase, and free energy calculations \cite{Smi05a,Smi05b}
and MD simulations \cite{Vrb05} support this fact, the observations for the direction of
preferential motion of positive/negative ions at the ice-water interface can shed light on
the Workman-Reynolds effect \cite{Wor50}.
We observed in MD simulations of the stable ice-water
interface (the mass-density profile for the equilibrium relaxed basal face of ice $I_h$ in
contact with water simulated with the SPC/E model \cite{Jor83} at $T=225$~K is shown in
figure~\ref{profile}) the same tendencies in the preferential direction of motion for simple Li$^{+}$ and F$^{-}$ ions (see figure~\ref{ions}) as were observed earlier for the larger
in size Na$^{+}$ and Cl$^{-}$ solute ions at the liquid side of ice-water interface \cite{Hay05}.
It is seen from figure~\ref{ions}, that the negative F$^{-}$ ion moves on the liquid side of the ice-water interface and is heading towars the bulk ice region, while the positive Li$^{+}$ ion shows a very similar behaviour with the Na$^{+}$ \cite{Hay05} having short-time trapping in cages on the liquid side of the interface, although with the long-time tendency of motion towards the water bulk region.

\begin{figure}[!b]
\centerline{
\includegraphics[width=0.5\textwidth]{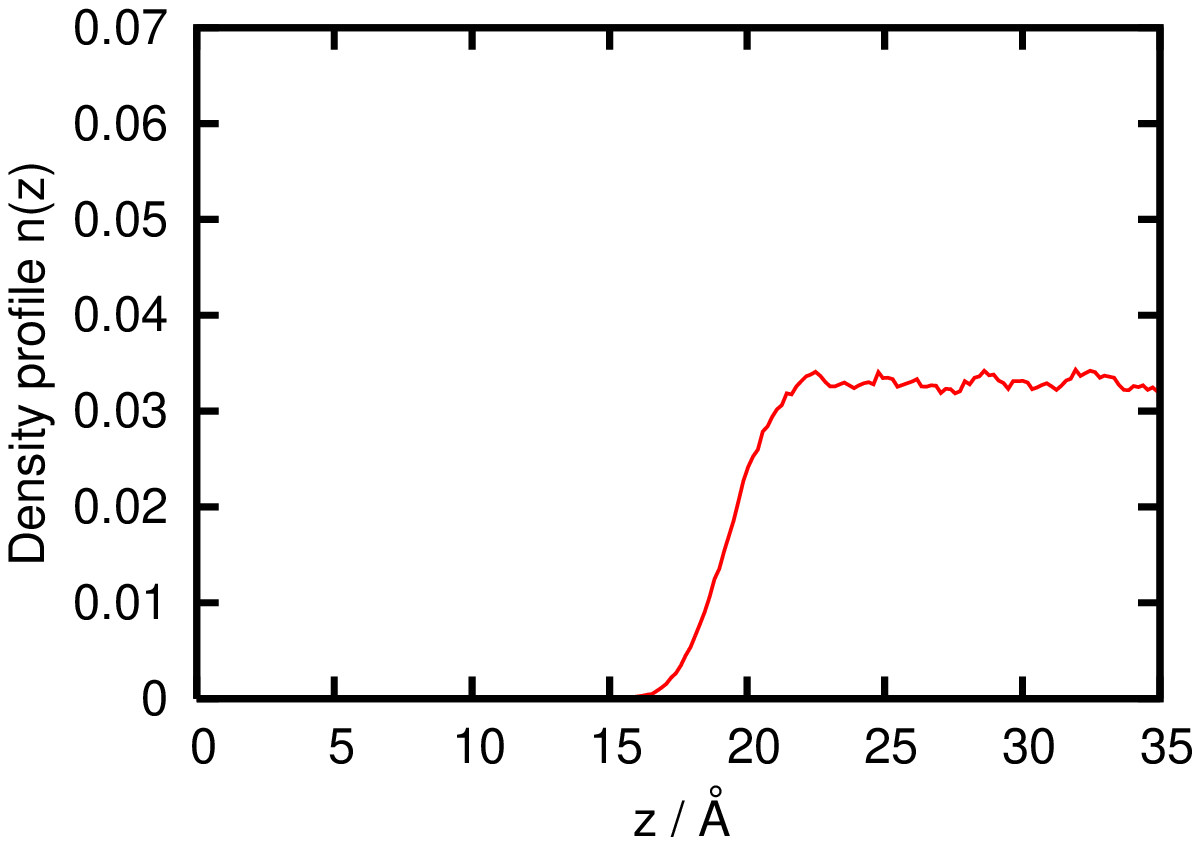}
\includegraphics[width=0.5\textwidth]{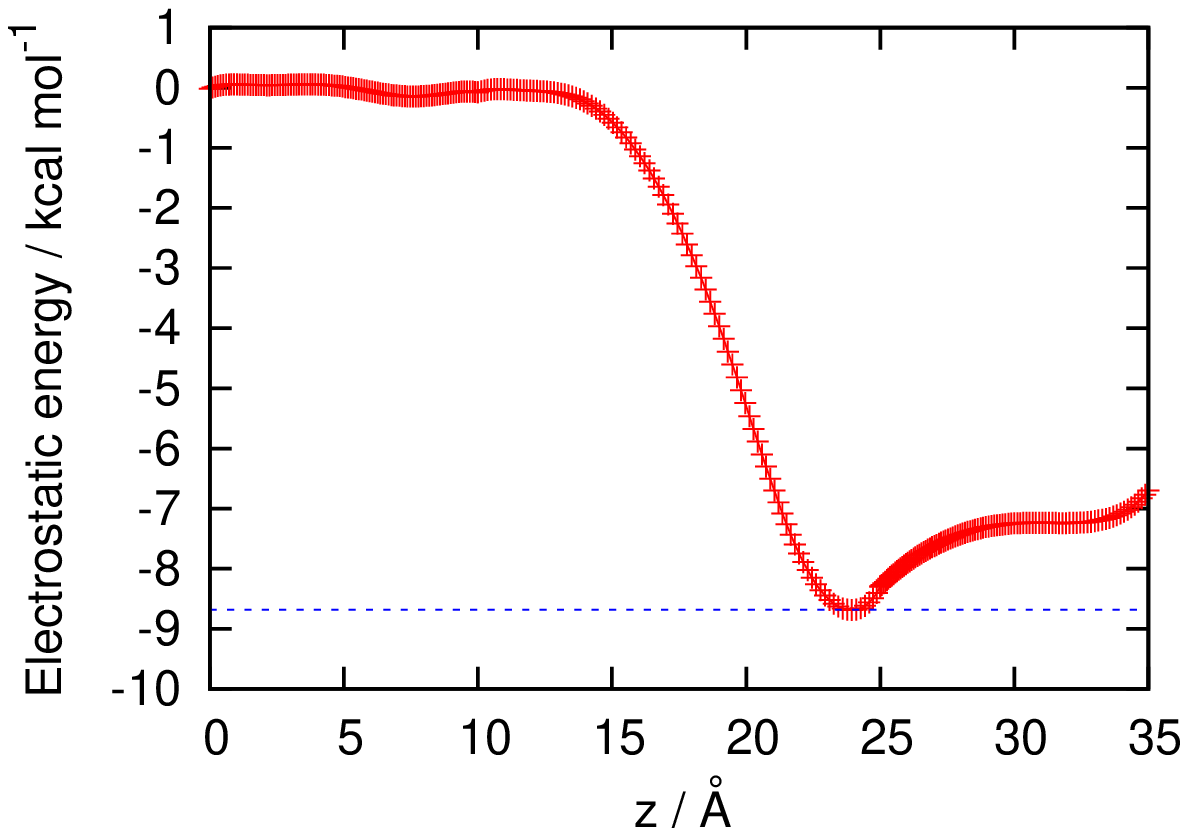}
}
\caption{(Color online) Number density profile and profile of electrostatic potential
for the water-vapor interface of SPC/E model at 298~K.
} \label{LG}
\end{figure}

Before calculating the profile of electrostatic potential across the ice-water interface
we tested our methodology on a simpler case of SPC/E water-vapor interface.
In figure~\ref{LG} we show the results for the density and electrostatic
potential profile across the water-vapor interface. The interface is
located between $z\approx 16$~{\AA} and $z\approx 23$~{\AA}, and namely in
this region one observes a strong drop of the electrostatic potential,
which shows a minimum nearly at the liquid side of the interface. In the literature
there were reported calculations of the electrostatic potentials for the
water-vapor interface of TIP4P water molecules \cite{Wil88,Wil89} and very similar (to
our case of SPC/E model)
profile of electrostatic potential was obtained. The minimum in profile of electrostatic
potential was observed at the distance $\sim 2.5$~{\AA} in the interface with respect to the
Gibbs dividing plane, and the drop of potential between the vapor and minimum in the
electrostatic potential in \cite{Wil89} was $\sim -0.4$~eV. In our case of the SPC/E
water-vapor interface, we observed a minimum in the obtained profile of the electrostatatic
potential also approximately at a distance 3~{\AA} from the Gibbs dividing surface of the
water-vapor density profile, and the drop in electrostatic potential was $\sim8.68$~kcal/mol (figure~\ref{LG}), which corresponds to $-376.4$~meV, and is in good agreement for the estimate
for TIP4P model \cite{Wil88}, keeping in mind that the two water models (SPC/E and TIP4P)
result in slightly different values of dielectric permittivity. Hence, our proposed
methodology of estimation of the profile
of electrostatic potential for the case of water-vapor interface gives a very reasonable
agreement with the calculations by other methodology.

\begin{figure}[!t]
\centerline{
\includegraphics[width=0.5\textwidth]{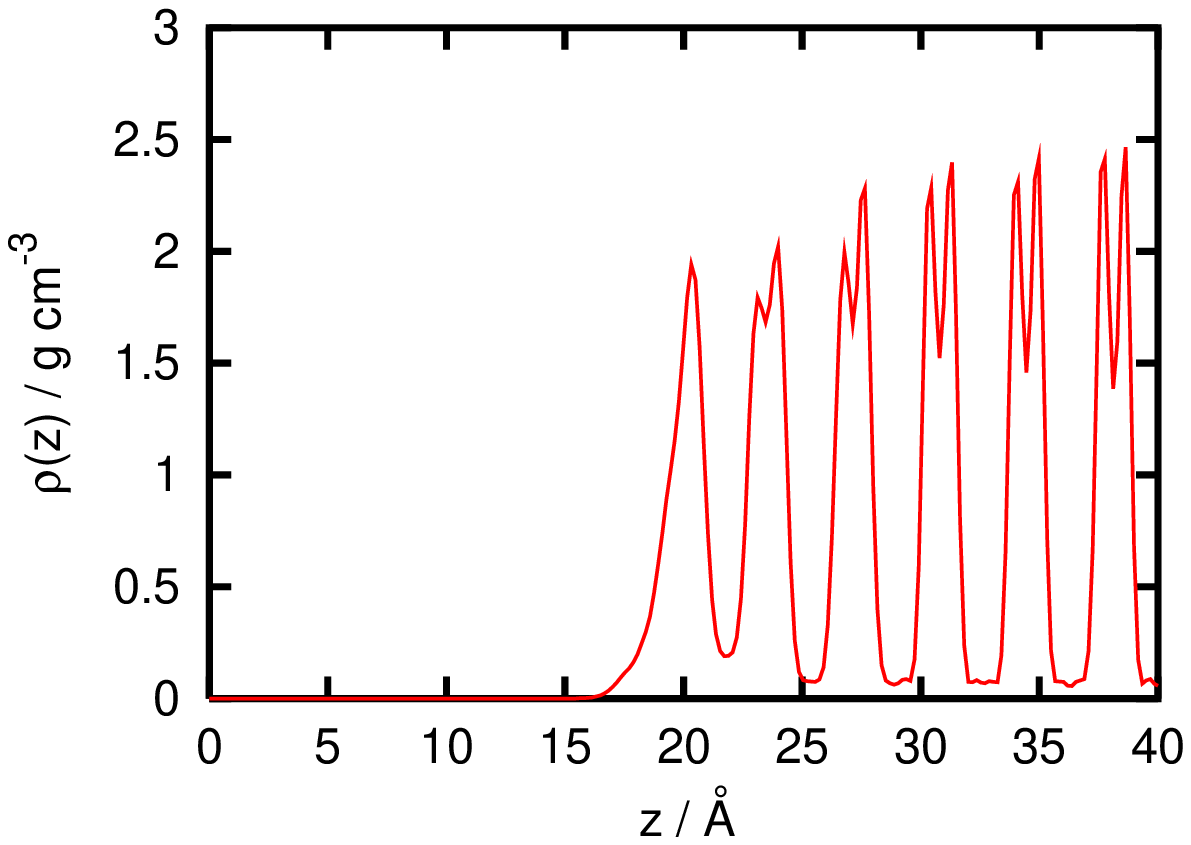}
\includegraphics[width=0.5\textwidth]{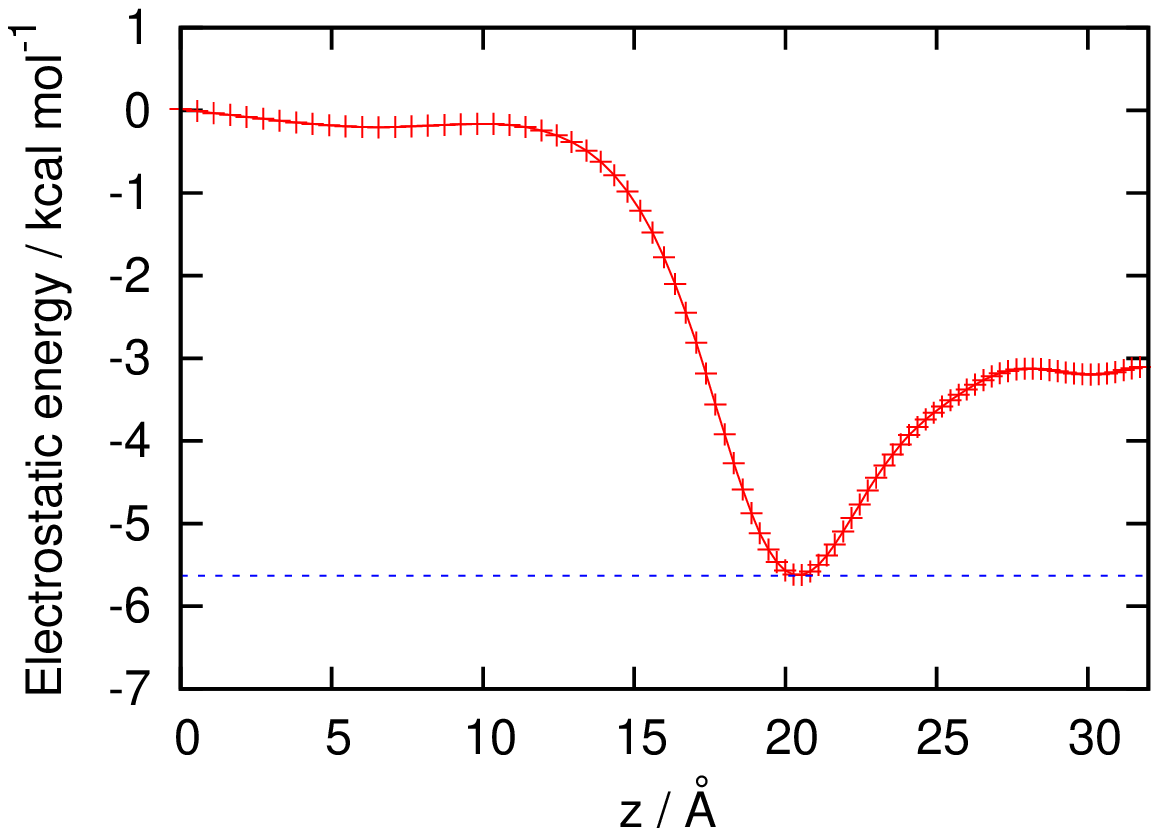}
}
\caption{(Color online) Mass-density profile and profile of electrostatic potential
for the ice (basal) surface at temperature 225~K.
} \label{ice}
\end{figure}

For the case of the pure ice surface, we used the same methodology of calculation of the
profile of electrostatic potential. The obtained density profile and electrostatic potential are
shown in figure~\ref{ice}. Qualitatively, the profile of eletrostatic potential for the ice
surface is similar to the water-vapor interface, although the surface potential is smaller
than for the water-vapor interface and is $\sim 5.63$~kcal/mol. There is a well-pronounced minimum
in the electrostatic potential located in the top smeared-out atomistic layer which actually
corresponds to the liquid-like layer on the ice surface. The MD simulations of the solute ions at
the ice surface \cite{Bry04b} gave evidence of localization of the positive solute ions in the
liquid-like layer, while the negative ions were trying to penetrate deeper in the interface creating
more defects and practically destroying the remaining order in the top surface layers.
In the literature, there is not much information on the surface potential of pure ice surface.
We were able to find some experimental estimates for the pure ice surface potential from
measurements with gold electrodes at temperatures between $0^\circ$C and $-15^\circ$C  \cite{Car80}. The ice surface
potential with respect to gold electrode was reported to be in the range $\sim 185-195$~meV, which,
however, upon riming changed the sign and became $\sim -215$~meV. Our MD calculations for the
pure ice surface
potential (with an existing liquid-like layer) give the value of $\sim -244$~meV. It is obvious that the
presence of the liquid-like layer on the surface of ice at melting temperature (at which
the MD simulations were performed) should give a comparable value and the same sign of the surface
potential as were obtained for the water-vapor interface.

\begin{figure}[!b]
\centerline{
\includegraphics[width=0.55\textwidth]{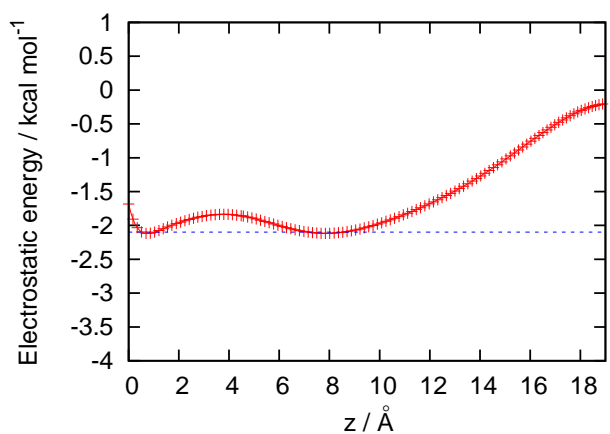}
}
\caption{(Color online) Profile of electrostatic potential
for the ice (basal face)-water surface of the SPC/E model, which corresponds to the
mass-density profile shown in figure~\ref{profile}.
} \label{LS}
\end{figure}

Figure~\ref{LS} reports the profile of electrostatic potential, that corresponds
to the ice-water interface with the mass-density profile shown in figure~\ref{profile}.
Our calculations show an increase of the electrostatic potential from the water
to the ice phase with approximately 2.1~kcal/mol ($\sim 91$~meV) net difference. This is in agreement
with our observations of the tendencies of the preferential motion of the positive
solute ion towards liquid phase and of the negative ion towards the bulk ice. Note
that the ice-water interface has the smallest change in the profile of electrostatic
potential in comparison with the water-vapor interface and ice surface. It is obvious that in the crystal bulk region the profile
of electrostatic potential should have the periodicity of the charge-density profile. We did not extend our calculations
into the ice bulk region for the ice-water and ice-vapor interfaces. In order to recover the periodicity of the
electrostatic potential in ice bulk region one has to use in
our methodology much smaller step in constrained $z$-position
for the trial particle.

\section{Conclusions}
\label{sec4}

In MD simulations of Li$^{+}$ and F$^{-}$ ions at the ice-water interface
we have observed different tendencies in the preferential direction of motion which were the same as in the earlier
reported behavior of larger in size ions Na$^{+}$ and Cl$^{-}$ in MD simulations \cite{Hay05}.
In both cases, the positive ions had a preferential direction of motion towards the bulk water
region, while the negative ions showed an opposite direction and were trying to penetrate deeper
in the interface. These observations are in agreement with the Workman-Reynolds effect of emerging
freezing potential at the growing ice in the aqueous solutions, which reflects the charge separation
with the excess of positive ions in the liquid phase.

In order to get insight into electrostatic effects which occur at different interfaces, we
calculated from MD simulations the profiles of electrostatic potential for the pure ice-water,
water-vapor
interfaces and pure ice surface. We have proposed a new methodology of calculating the profiles of electrostatic
potentials by making use of a trial particle which moves in a plane with a fixed
$z$-coordinate and due to soft-core repulsion avoids the positions of point charges. Calculations
of a single-particle electrostatic energy by Ewald method at the instantaneous position of the trial
particle is connected with the electrostatic potential at this point, which after the average over
different configurations results in the profile of electrostatic potential. Our check performed
for the water-vapor interface showed good agreement with the profile of electrostatic potential for
the water-vapor interface simulated with TIP4P model \cite{Wil88}.

Our results indicate that among the three interfaces, the pure water-vapor interface has the largest decrease of the profile of electrostatic potential in the liquid phase, while the ice-water interface
has the smallest difference. The obtained profile of electrostatic potential for the pure ice-water
interface decreases towards the bulk water region which can explain the observed difference in the
preferential direction of motion of the positive/negative solute ions observed in MD simulations.


%

\ukrainianpart

\title{Профілі електростатичного потенціалу для границь розділу вода-пара,
лід-пара та лід-вода}
\author{Т. Брик\refaddr{label1,label2}, А.Д.Дж. Геймет\refaddr{label3}}
\addresses{
\addr{label1} Інститут фізики конденсованих систем НАН України,
вул. І.~Свєнціцького,~1, 79011 Львів, Україна
\addr{label2} Інститут прикладної математики та фундаментальних наук,
Національний Університет ``Львівська Політехніка'', 79013 Львів, Україна
\addr{label3} Інститут океанографії Скріппса, Університет Каліфорнії
Сан-Дієго, Сан-Дієго, Каліфорнія 92093-0210, США
}

\makeukrtitle

\begin{abstract}
\tolerance=3000%
Границі розділу лід-вода, вода-пара та поверхня льоду були досліджені комп'ютерним моделюванням
методом молекулярної динаміки з SPC/E моделлю молекул води маючи на меті визначити профілі
електростатичного потенціалу через границі розділу фаз. Ми запропонували методологію для
розрахунку профілів електростатичного потенціалу з використанням пробної частинки. Запропонований
підхід показав добре узгодження з профілем електростатичного потенціалу для границі розділу
вода-пара в моделі води, розрахованого іншою методикою. Отриманий профіль електростатичного
потенціалу для чистої границі розділу лід-вода зменшується в сторону рідкої фази, що
узгоджується з моделюванням переважаючого напрямку руху домішкових іонів Li$^{+}$ та F$^{-}$
на рідинній стороні границі розділу лід-вода. Отримані результати обговорюються з точки
зору їх відношення до ефекту Воркмана-Рейнольдса.

\keywords границя розділу лід-вода, поверхня льоду, границя розділу вода-пара,
 домішкові іони, профіль електростатичного
потенціалу, ефект Воркмана-Рейнольдса, моделювання методом молекулярної динаміки

\end{abstract}

\end{document}